# First-principles prediction of mechanical and bonding characteristics of new $T_2$ superconductor $Ta_5GeB_2$

**M. A. Hadi**[*,1], **M. T. Nasir**[2], **M. Roknuzzaman**[3], **M. A. Rayhan**[2], **S. H. Naqib**[1] **and A. K. M. A. Islam**[1,4]

[1] Department of Physics, University of Rajshahi, Rajshahi-6205, Bangladesh
[2] Department of Arts & Science, Bangladesh Army University of Science & Technology, Saidpur-5310, Nilphamari, Bangladesh
[3] Department of Physics, Jessore University of Science and Technology, Jessore-7408, Bangladesh
[4] International Islamic University Chittagong, 154/A College Road, Chittagong-4203, Bangladesh

* Corresponding author: E-mail hadipab@gmail.com, Phone: +88 01716200042

**Abstract**

In the present paper, DFT (Density Functional Theory) based first-principles methods are applied to investigate the mechanical and bonding properties of newly synthesized $T_2$ phase superconductor $Ta_5GeB_2$ for the first time. The calculated lattice constants are in reasonable agreement with the experiment. The elastic constants ($C_{ij}$), bulk modulus ($B$), shear modulus ($G$), Young's modulus ($Y$), Poisson ratio ($v$), Pugh ratio ($G/B$), and elastic anisotropy factor $A$ of $Ta_5GeB_2$ are calculated and used to explore the mechanical behavior of the compound. To give an explanation of the bonding nature of this new ternary tetragonal system, the band structure, density of states, and Mulliken atomic population are investigated. The estimated Debye temperature and Vickers hardness are also used to justify both the mechanical and bonding properties of $Ta_5GeB_2$.

## 1 Introduction

Very recently, Corrêa et al. [1] synthesized a new compound $Ta_5GeB_2$ belonging to the tetragonal $T_2$ phase with $Cr_5B_3$ prototype structure that is stabilized by substituting Ge for B at the 8h Wyckoff position. In addition to magnetization, electrical resistivity and heat capacity measurements, they also reported on the bulk superconductivity of this compound with a transition temperature $T_C$ ~3.8 K. The $T_2$ phases are labeled as "5-3" metal-metalloid compounds which fall in the large group of metal based systems spanning from the alkaline-earth metal to the late transition metals. These phases comprise more than 40 compounds of the general formula $M_5X_3$ (M = metal, X = semi-metal or non-metal), which form tetragonal crystal structure, usually known as the $Cr_5B_3$ type [2]. The $T_2$ structure also retains a rather high-coordination number of metal–metal atomic bonds to maintain a fairly close-packed structure [3,4]. In addition, the $T_2$ crystal structure retains a body-centered symmetry similar to refractory metals. The majority of these materials crystallize into the space group I4/*mcm* (No. 140). Some lower symmetry distorted variants are also known. In accordance with the occupied crystallographic sites in I4/*mcm*, the general composition $M_5X_3$ can be rewritten as $M(l)M(2)_4X(l)X(2)_2$. There are significant differences in the cell dimensions and site parameters among some compounds, though most of them crystallize in the same space group. In fact, the $Cr_5B_3$ type is divided into two isopointal subfamilies, namely major and minor subfamilies [5,6]. All the members of the major subfamily known to date are typified by a *c/a* ratio of about 1.85 and by the formation of $X(2)_2$ dumbbells. The shortest inter-atomic distance in the major subfamily occurs between two X(2) atoms. This distance is due to a covalent single bond. The compound $Cr_5B_3$ falls in this subfamily. This subfamily is also known as $Cr_5B_3$ subfamily. The minor subfamily, the so-called $In_5Bi_3$ subfamily, includes three binary compounds as its members. The members $In_5Bi_3$ and $Tl_5Te_3$ [5,7-14] of this subfamily are well known. In these compounds, the *c/a* ratio is about 1.45 and no dumbbells are seen to occur. Shorter interatomic distances are observed between M(1) and X(1) atoms in a linear chain along [00l] in this subfamily.

$T_2$ phase tenders a collection of fascinating properties, such as high-melting temperature [15], oxidation resistance [16-18], and advantageous high-temperature mechanical properties [19-21]. As the focal point of the microstructure designs with the $T_2$ ternary phase, the primary basis of the alloying nature in this phase together with the communal solid solution of a wide range of transition metals has been formed in accordance with the governing geometric and electronic aspects. In a multi-phase alloy the materials show evidence of high-temperature creep strength [22,23] and ambient temperature flexural strength [24]. The remarkable properties of the multi-phase alloys have raised the interest in the monolithic $T_2$ phase. For instance, the recent data from the room temperature Vickers indentation tests point out that the hardness and the fracture toughness for the $T_2$ phase are about 30% higher than those of the $T_1$ phase [25]. In addition, the accumulation of congenital vacancies has been exposed to perform an important task in development of dislocation and precipitation reactions in the $T_2$ phase that have direct impact on high-temperature structural performance. The typically slow diffusion rates inside the $T_2$ phase have also been quantified and used to the materials processing approaches.

In the present study, a detailed theoretical investigation of the ground state mechanical and bonding properties of new $T_2$ phase superconductor $Ta_5GeB_2$ has been done within the plane wave pseudopotential technique. The mechanical behaviors are analyzed in terms of the single crystal elastic constants and polycrystalline elastic moduli. The bonding characteristics are described by means of electronic band structure, electron energy density of states and Mulliken atomic populations.

## 2 Methods of calculations

The DFT [26,27] calculations are carried out on the new $T_2$ superconductor $Ta_5GeB_2$ using the CASTEP (Cambridge Serial Total Energy Package) code [28]. The computations are performed in a unit cell containing four formula units with 32 atoms (20 Ta, 4 Ge, and 8 B). The GGA-PBE exchange-correlation [29] is applied with the plane-wave pseudopotential available in the above mentioned code. The Vanderbilt-type ultrasoft pseudopotential [30] is used to take care of electron-ion interactions. For sampling the first Brillouin zone, a k-point grid of 7 × 7 × 4 mesh according to Monkhorst-Pack scheme [31] is set for all calculations providing a spacing of 0.02 Å$^{-1}$. An energy cutoff of 500 eV is chosen in order to limit the number of plane-waves in the expansion. The BFGS (Broyden-Fletcher-Goldfarb-Shanno) minimization method [32] is utilized for searching the ground state of crystal. All calculations are accomplished under zero pressure, allowing all atomic sites, lattice constants and angles to fully relax. To optimize the system geometry, the chosen tolerances are: total energy difference per atom within 5 × 10$^{-6}$ eV, maximum ionic Hellmann-Feynman force within 0.01 eV/Å, maximum stress under 0.02 GPa, and maximum ionic displacement under 5 × 10$^{-4}$ Å. The elastic constants, band structure, density of states, and Mulliken atomic population of the new compound are extracted from the fully geometry optimized state.

## 3 Results and Discussions

### 3.1 Structural properties

The $T_2$ phase $Ta_5GeB_2$ has been found to crystallize in the tetragonal structure $D8_1$ with space group I4/*mcm* (No. 140). Here, the atom Ta is at 4c (0, 0, 0) and 16l ($x$, $x$ +1/2, $z$) Wyckoff positions. At the same time, the p elements Ge and B are positioned at 4a (0, 0, 1/4) and 8h ($x$, $x$ +1/2, 0) Wyckoff positions, respectively. $Ta_5GeB_2$ structure [Fig.1] is formed by alternate layers. The first one is at ($z$ = 0) including both the metal Ta and the p element B, the second ($z$ = 1/8) only the transition metal, the third ($z$ = 1/4) only the p element Ge, the fourth ($z$ = 3/8) only metal Ta, the fifth ($z$ = 1/2) comprising both the metal Ta and the p element B. The other layers are due to the tetragonal centered symmetry. For transition metal and p element alloys, the $D8_1$ structure shows stability when the concentration of the valence electron are from 4.3 to 5.5 [33]. With a valence electron concentration of 4.375, the new $T_2$ phase $Ta_5GeB_2$ satisfies the above stability condition. Optimized lattice constants and $x$ parameters of $Ta_5GeB_2$ are listed in Table 1 along with those obtained for some other isostructural compounds. The computed lattice constants deviate less than 1.8 % of the experimental values. The evaluated lattice constants are slightly larger than the measured values, which is a general trend inherent to GGA calculations.

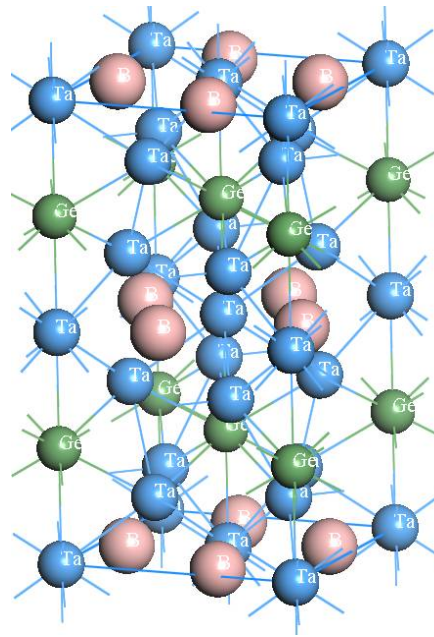

**Figure 1** Crystal structure of new $T_2$ superconductor $Ta_5GeB_2$.

**Table 1**

Structural parameters for $Ta_5GeB_2$ with the different $T_2$ phases of $D8_l$ structure, space group $I4/mcm$, transition metal in 4c and 16l, Si/Ge and B in 4a and 8h.

| $T_2$ phases | $a$ (Å) | $c$ (Å) | $c/a$ | $x_{8h}$ | $x_{16l}$ | $z_{16l}$ | References |
|---|---|---|---|---|---|---|---|
| $Ta_5GeB_2$ | 6.334 | 11.785 | 1.861 | 0.3827 | 0.1663 | 0.1381 | Calc. [This] |
|  | 6.239 | 11.578 | 1.856 |  |  |  | Expt. [1] |
| $Nb_5SiB_2$ | 6.245 | 11.655 | 1.866 | 0.6123 | 0.1695 | 0.1375 | Calc. [34] |
| $Mo_5SiB_2$ | 6.037 | 11.114 | 1.841 |  |  |  | Calc. [35] |
|  | 6.03 | 11.0 | 1.824 |  |  |  | Calc. [36] |
|  | 6.013 | 11.0485 | 1.837 | 0.375 | 0.1653 | 0.1388 | Expt. [3] |
|  | 6.0272 | 11.0671 | 1.836 | 0.3784 | 0.1641 | 0.1398 | Expt. [37] |
|  | 6.001 | 11.022 | 1.837 |  |  |  | Expt. [38] |

### 3.2 Elastic properties

The elastic constants are computed within the CASTEP module from the first-principles method by using a set of uniform deformations of a finite value. The resulting stresses are then computed by optimizing the internal degrees of freedoms [39]. The Voigt-Reuss-Hill approximations [40-42] are used to calculate the polycrystalline bulk modulus ($B$) and shear modulus ($G$). The equations, $Y = (9GB)/(3B + G)$ and $v = (3B – 2G)/(6B + 2G)$ are also used to evaluate the Young's modulus $Y$ and Poisson ratio $v$, respectively.

The six elastic constants ($C_{11}$, $C_{12}$, $C_{13}$, $C_{33}$, $C_{44}$ and $C_{66}$) and elastic moduli ($B$, $G$, $Y$, $B/G$ and $v$) of tetragonal crystal system $Ta_5GeB_2$ are calculated and listed in the Table 2. The newly synthesized $T_2$ phase satisfies the mechanical stability conditions for tetragonal crystal [43]: $C_{11} > 0$, $C_{33} > 0$, $C_{44} > 0$, $C_{66} > 0$, $C_{11} – C_{12} > 0$, $C_{11} + C_{33} – 2 C_{13} > 0$, $2(C_{11} + C_{12}) + C_{33} + 4 C_{13} > 0$. The elastic constants $C_{11}$ and $C_{33}$ describe the linear compression resistance along the directions $a$ and $c$, respectively. As can be seen that the calculated elastic constants $C_{11}$ and $C_{33}$ are very large compared to other elastic constants, indicating that the $Ta_5GeB_2$ system is very incompressible under uniaxial stress along the directions $a$ and $c$. The elastic constant $C_{33}$ is much larger than $C_{11}$, meaning that the incompressibility along the direction c is much higher. Indeed, the bonds collateral with the $c$-axis show a dominating effect on $C_{33}$ making it much larger than $C_{11}$. Because of $C_{11}+C_{12} > C_{33}$, the bonding in the (001) plane is more rigid elastically than that along the c-axis as well as the elastic tensile modulus is higher on the (001) plane than that along the c-axis. As $C_{44}$ is an important parameter to describe the indentation hardness [44], it is expected that the hardness of $Ta_5GeB_2$ should be similar to that of $Mo_5SiB_2$.

**Table 2**

Single crystal elastic constants $C_{ij}$, polycrystalline bulk modulus $B$, shear modulus $G$, and Young's modulus $Y$, in GPa, Pugh's ratio $G/B$, and Poisson ratio $v$ of $Ta_5GeB_2$ in comparison with other $T_2$ phase.

| Crystals | $C_{11}$ | $C_{12}$ | $C_{13}$ | $C_{33}$ | $C_{44}$ | $C_{66}$ | $B$ | $G$ | $Y$ | $B/G$ | $v$ | References |
|---|---|---|---|---|---|---|---|---|---|---|---|---|
| $Ta_5GeB_2$ | 340 | 124 | 137 | 391 | 169 | 137 | 230 | 152 | 374 | 1.51 | 0.23 | [This] |
| $Mo_5SiB_2$ | 483 | 154 | 188 | 419 | 179 | 127 |  |  |  |  |  | [36] |
|  | 479 | 174 | 203 | 390 | 163 | 138 |  |  |  |  |  | [35] |
|  | 480 | 166 | 197 | 415 | 174 | 143 | 277 | 151 | 383 | 1.83 | 0.27 | [45] |
| $Ta_5GeC_4$ | 460 | 162 | 197 | 384 | 166 | 148 | 268 | 143 | 365 | 1.87 | 0.27 | [49] |

In the present calculations, $C_{44} > C_{66}$, which insures that the new $T_2$ superconductor $Ta_5GeB_2$ is $Cr_5B_3$-prototype phase and in which the [100](010) shear is easier than the [100](001) shear. The shear anisotropy factor $A$, [= $2C_{66}/(C_{11}$ - $C_{12}$)] is found to be 1.27. This indicates that for the (001) plane of Ta$_5$GeB$_2$ the shear elastic properties are strongly dependent on the shear directions.

Cauchy pressure, Pugh's ratio ($B/G$) and Poisson ratio ($v$) are regarded as a measure to predict the failure mode, i.e., ductile versus brittle nature, of materials. Materials which easily change their volumes are brittle and materials which can be easily distorted are ductile. The Cauchy pressure ($C_{12} - C_{44}$) is considered to serve as an indication of ductility/brittleness of materials. The material is likely to be ductile (brittle) when the pressure is positive (negative). A different index of the ductility/brittleness is the Pugh ratio ($B/G$). The high or low ratio is linked to the ductile or brittle nature. To take apart the ductile materials from brittle ones, the critical value is found to be 1.75. Frantsevich et al. [46] also indicated the separation of the ductility from brittleness of materials on the basis of Poisson ratio. Frantsevich rule suggests $v \sim 0.26$ as the border line which separates the brittle from ductile materials. If the Poisson ratio is greater than 0.26 then the material will be ductile otherwise the material will be brittle. The Cauchy pressure ($C_{12} - C_{44}$) of Ta$_5$GeB$_2$ is negative, its Pugh ratio $B/G$ is 1.51 which is less than 1.75 and its Poisson ratio $v$ is 0.23 which is less than 0.26. So, the compound Ta$_5$GeB$_2$ should be brittle in nature.

Bulk modulus $B$ assesses the resistance to fracture and shear modulus $G$ estimates the resistance to plastic deformation of polycrystalline materials. The bulk modulus bears a little connection with hardness, as is well known from dislocation theory [47]. On the other hand, a better correlation is observed between hardness and shear modulus. Indeed, the hardness is more sensitive to the shear modulus than the bulk modulus. Therefore, the hardness of Ta$_5$GeB$_2$ and Mo$_5$SiB$_2$ are expected to be similar.

The Young's modulus $Y$ evaluates the resistance against longitudinal tensions. In addition, the Young's modulus has influence on the thermal shock resistance of a material, as the critical thermal shock coefficient $R$ is inversely proportional to the Young's modulus $Y$ [48]. The larger the $R$ value, the better the thermal shock resistance. The thermal shock resistance is an essential indicator for thermal barrier coating (TBC) materials selection. The small $Y$ value indicates that Ta$_5$GeB$_2$ is more resistant to thermal shock than Mo$_5$SiB$_2$. All of its elastic properties are also compared with a MAX phase compound Ta$_5$GeC$_4$ [49]. In this system the only difference is B against C. Surprisingly, excepting C$_{11}$, all of its elastic properties match quite well with the results obtained for the new T$_2$ phase superconductor, Ta$_5$GeB$_2$.

### 3.3 Debye temperature

Debye temperature $\theta_D$ is associated with various important properties of materials. In fact, a higher Debye temperature implies a higher phonon thermal conductivity. The excitation due to vibration at low temperatures arises solely from acoustic vibrations. In order to estimate the magnitude of Debye temperature from the mean sound velocity, we used the following equation [50]:

$$\theta_D = h/k_B[(3n/4\pi)N_A\rho/M]^{1/3}v_m$$

where $M$ is the molecular mass, $n$ is the number of atoms per formula unit, and $\rho$ is the mass density; $h$ is the Plank's constant, $k_B$ is the Boltzmann constant, and $N_A$ is the Avogadro's number. For polycrystalline material the average sound velocity is expressed as [50]:

$$v_m = [1/3(1/v_l^3 + 2/v_t^3)]^{-1/3}$$

where $v_l$ and $v_t$ represent the longitudinal and transverse sound velocities in an isotropic material. These can be determined in terms of bulk modulus $B$ and shear modulus $G$:

$$v_l = [(3B + 4G)/3\rho]^{1/2} \quad \text{and} \quad v_t = [G/\rho]^{1/2}.$$

The calculated value of Debye temperature is listed in Table 3 along with measured value as well as with the values for other T$_2$ phase. The calculated result is deviated by 7.76% from the experimental value. This deviation is expected because the calculation is done on perfect crystal while the measured values are dependent on the purity of the sample, in which impurities, defects, and grain boundaries may be present. The theoretical result predicts that the new compound has a relatively low $\theta_D$, indicating that it possesses a rather flexible lattice and hence low thermal conductivity.

**Table 3**

Calculated mass density ($\rho$ in gm/cm$^3$), longitudinal, transverse and sound velocities ($v_l$, $v_t$, and $v_m$ in km/s) and Debye temperature ($\theta_D$ in K) of $Ta_5GeB_2$ in comparison with $Mo_5SiB_2$.

| $T_2$ phase | $\rho$ | $v_t$ | $v_l$ | $v_m$ | $\theta_D$ | References |
|---|---|---|---|---|---|---|
| $Ta_5GeB_2$ | 14.7 | 3.216 | 5.425 | 3.561 | 375 | Calc. [This] |
|  |  |  |  |  | 348 | Expt. [1] |
| $Mo_5SiB_2$ |  |  |  |  | 592 | Calc. [45] |
|  |  |  |  |  | 515 | Expt. [38] |

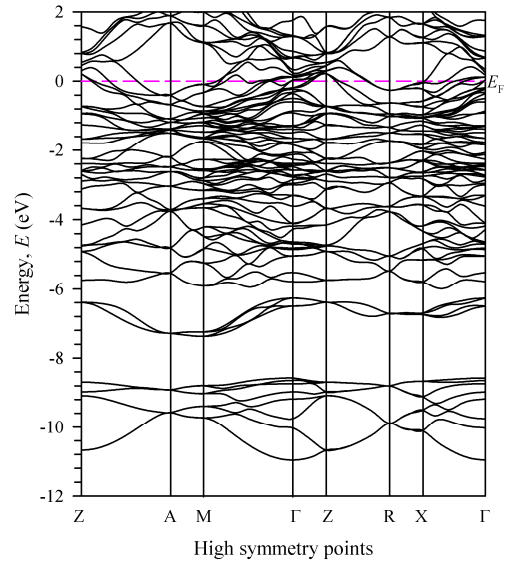

**Figure 2** Electronic band structure of $T_2$ phase $Ta_5GeB_2$.

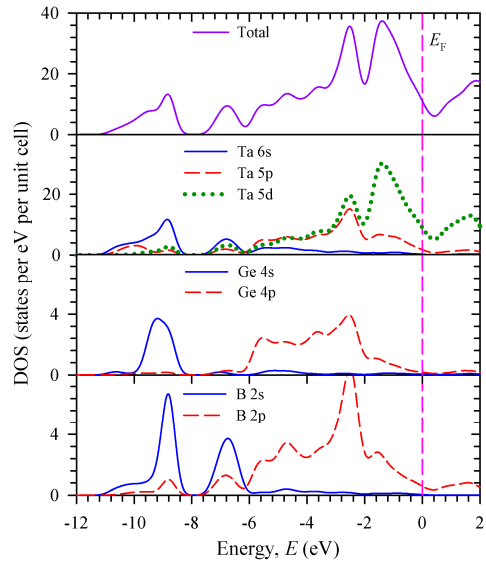

**Figure 3** Total and partial DOS of $Ta_5GeB_2$. The Fermi level $E_F$ is set at 0 eV.

### 3.4 Band structure and DOS

The electronic properties of new ternary compound Ta$_5$GeB$_2$ is investigated by calculating the band structure and total as well as partial DOSs. The calculated band structure at equilibrium lattice constants is drawn along high symmetrical directions in the first Brillouin zone [Fig. 2]. The Fermi level of new T$_2$ superconductor lies below the valence bands maximum near the Γ point. The new T$_2$ phase is observed to be metallic in nature since there is a overlapping of valence bands with conduction bands.

The nature of bonding of solids may be suitably predicted through partial DOS analysis. To describe the electronic structure and bonding nature, the total and partial DOSs are shown in Fig. 2b. The large value of DOS at $E_F$ assures once again the metallic nature of new T$_2$ superconductor Ta$_5$GeB$_2$. The Fermi level lies to the left of a deep valley named pseudogap $E_P$, which indicates the stability of the structure. Electrons situated at levels above $E_P$ become delocalized and the material becomes metalized. Mainly Ta 5d states contribute to the DOS at the Fermi level with small contribution from p states of Ta and B. The calculated DOS at the Fermi level is found to be 11.4 states/cell/eV. This large value indicates the high metallic conductivity of the new compound.

The valence bands lying in the lowest energy range of -11.1 to -8.2 eV originates mainly from Ta 6s states with admixture from Ge 4s and B 2s states. These valence bands are seen to be separated by a narrow forbidden gap of ~0.6 eV form the higher valence bands situated in the energy range from -7.6 eV to $E_F$. The wide higher valence band possesses several distinct peaks. It is seen that the left peak structure between -7.6 and -6.2 eV is formed almost entirely by Ta 6s and B 2s states. The next structures in the total DOS situated between -6.2 and -3.3 eV are mainly due to Ta 5p and 5d states as well as Ge 4p and B 2p states. The second highest peak, between -3.3 and -2.0 eV consists of Ta 5d and 5p states. The highest peak between -2.0 eV and $E_F$ is formed from the strong hybridization of p and d states of Ta with p states of Ge and B. It is obvious that within the range from -7.6 eV to $E_F$, there occurs a covalent interaction between the constituting atoms. This is due to the reason that states are really degenerate with respect to angular momentum and lattice site. Further, some degree of ionic character is expected simply for the difference in the value of electronegativity among the comprising elements.

### 3.5 Mulliken atomic populations

The Mulliken population analysis can facilitate to assign the electrons in several fractional methods among the various parts of bonds and overlap population has correlations with covalency or ionicity of bonding and bond strength. Population analysis in CASTEP is carried out using a projection of the plane wave states onto a localized basis by means of a technique developed by Sanchez-Portal et al. [51]. Population analysis of the resulting projected states is then accomplished by using the Mulliken formalism [52]. This method is widely used in the analysis of electronic structure calculations performed with Linear Combination of Atomic Orbitals (LCAO) basis sets. The most essential quantities in relation to atomic bond population calculations are the effective charge and the bond order values between pairs of bonding atoms using minimal basis within the Mulliken scheme [52,53] as follows:

$$Q_\alpha^* = \sum_{i,\alpha} \sum_{n\ occ} \sum_{j,\beta} C_{i\alpha}^{*n} C_{j\beta}^n S_{i\alpha,j\beta}$$

$$\rho_{\alpha\beta} = \sum_{n\ occ} \sum_{i,j} C_{i\alpha}^{*n} C_{j\beta}^n S_{i\alpha,j\beta}$$

where $i, j$ stand for the orbital quantum numbers and $n$ is the band index, $C_{i\alpha}^{*n}$ as well as $C_{j\beta}^n$ are the eigenvector coefficients of the wave function and $S_{i\alpha,j\beta}$ is the overlap matrix between atoms $\alpha$ and $\beta$.

The effective valence charge and Mulliken atomic population constantly assist to realize the bonding nature in solids. The difference between the formal ionic charge and Mulliken charge on the anion species in the crystal estimates the effective valence and verifies the ascendancy of covalent and ionic bonding. An ideal ionic bond is observed when the effective valence has zero value. On the contrary, an increasing level of covalency is seen if the effective valence carries a value greater than zero. The effective valence listed in the Table 4 would be a sign of the prominent covalency in bonding within the new ternary T$_2$ phase Ta$_5$GeB$_2$. The calculated bond overlap populations for the new ternary T$_2$ phase superconductor listed in Table 5. The overlap population of nearly zero value signifies that the interaction between the electronic populations of the two atoms is insignificant. A bond associated with a smallest Mulliken population is extremely weak and which plays insignificant role in the materials hardness. A low overlap population is a sign of a high degree of ionicity, whereas a high value implies a high degree of covalency in the chemical bond. The bonding and anti-bonding states arise due to positive and negative bond overlap populations, respectively. It is seen that the B–B bonds are more covalent than the Ta-Ta bonds.

**Table 4** Population analysis of Ta$_2$GeB$_2$.

| Species | Mulliken Atomic populations | | | | | Effective valence Charge (e) |
|---|---|---|---|---|---|---|
| | s | P | d | Total | Charge(e) | |
| B | 1.18 | 2.53 | 0.00 | 3.72 | − 0.72 | -- |
| Ge | − 0.75 | 3.00 | 0.00 | 2.25 | 1.75 | 2.25 |
| Ta 1 | 0.63 | 0.38 | 3.90 | 4.91 | 0.09 | 4.91 |
| Ta 2 | 0.64 | 0.60 | 3.86 | 5.10 | − 0.10 | 5.10 |

### 3.6 Theoretical Vickers hardness

Hardness is defined as the ability of a material to resist plastic deformation. The resistant force acted on per unit area takes part in estimating the hardness of a material. To evaluate the hardness of non-metallic materials a formula is developed with Mulliken population in first-principles method [54]. This method is not applicable for compounds with partial metallic bonding like T$_2$ phases because metallic bonding is delocalized and has no direct relation with hardness [55]. To calculate the hardness of metallic crystals, a correction for metallic bonding in the formula should be taken into account. Gou et al. [56] proposed a formula including such correction for the bond hardness of a crystal having partial metallic bonding as follows:

$$H_v^\mu = 740(P^\mu - P^{\mu\prime})(v_b^\mu)^{-5/3}$$

where $P^\mu$ represents the Mulliken overlap population of the $\mu$-type bond, $P^{\mu\prime}$ is symbolized for the metallic population and is calculated using the unit cell volume $V$ and the number of free electrons in a cell $n_{free} = \int_{E_P}^{E_F} N(E)dE$ as $P^{\mu\prime} = n_{free}/V$, and $v_b^\mu$ stands for the volume of a bond of $\mu$-type, which is evaluated from the bond length $d^\mu$ of type $\mu$ and the number of bonds $N_b^\nu$ of type ν per unit volume by $v_b^\mu = (d^\mu)^3/\sum_\nu[(d^\mu)^3 N_b^\nu]$. The hardness for the complex multiband crystals can be calculated as a geometric average of all bond harnesses as follows [57,58]:

$$H_V = [\prod^\mu (H_v^\mu)^{n^\mu}]^{1/\Sigma n^\mu}$$

where $n^\mu$ is the number of bond of type $\mu$ composing the actual multiband crystal. The obtained value of the Vickers hardness for the newly synthesized compound is listed in Table 5. The hardness value of 14.6 GPa for Ta$_5$GeB$_2$ is slightly less than the room temperature Vickers hardness of around 18 GPa for Mo$_5$SiB$_2$ [45]. It is, still, essential to note that the calculated Vickers hardness is only 15.2% that of diamond (96 GPa), which stays behind the hardest material known to date.

pss-Header will be provided by the publisher                                                                                                              9**Table 5**

Calculated bond and Vickers hardness $H_v^\mu$, $H_v$ (in GPa) of $Ta_5GeB_2$ along with bond number $n^\mu$, bond length $d^\mu$ (Å), bond volume $v_b^\mu$ (Å$^3$) and bond and metallic populations $P^\mu$, $P^{\mu'}$.

| Bond | $n^\mu$ | $d^\mu$ | $P^\mu$ | $P^{\mu'}$ | $v_b^\mu$ | $H_v^\mu$ | $H_v$ |
|---|---|---|---|---|---|---|---|
| B–B | 4 | 2.10100 | 0.40 | 0.1084 | 2.2027 | 57.87 | 14.6 |
| B–Ta | 32 | 2.44347 | 0.34 | 0.1084 | 3.4649 | 21.60 | |
|  | 16 | 2.53161 | 0.44 | 0.1084 | 3.8536 | 25.91 | |
|  | 16 | 2.53545 | 0.26 | 0.1084 | 3.8711 | 11.75 | |
| Ta–Ta | 32 | 2.86824 | 0.30 | 0.1084 | 5.6043 | 8.02 | |
|  | 8 | 2.97855 | 0.35 | 0.1084 | 6.2761 | 8.37 | |

## 4. Conclusion

In this paper, first-principles investigations have been carried out on the structural parameters, single crystal as well as polycrystalline elastic properties, Debye temperature, band structure, DOS, Mulliken atomic population and theoretical Vickers hardness of newly synthesized $Ta_5GeB_2$ for the first time. The calculated lattice parameters show good agreement with the measured values. The mechanical properties are estimated from the calculated elastic constants. Born criteria for tetragonal crystal are satisfied, suggesting that the new compound $Ta_5GeB_2$ is mechanically stable. Cauchy pressure, Pugh's ratio and Poisson ratio predict that $Ta_5GeB_2$ should be characterized as brittle material. The small value of $Y$ indicates that $Ta_5GeB_2$ should have good thermal shock resistance. The relatively low Debye temperature of 375 K should be a sign of flexible lattice as well as low thermal conductivity. The bonding nature of $Ta_5GeB_2$ may be explained as a mixture of metallic, covalent and ionic. In $Ta_5GeB_2$, the B–B bonds are more covalent than the Ta-Ta bonds. The calculated hardness value of 14.6 GPa for $Ta_5GeB_2$ is 15.2% of (96 GPa) for diamond, which is the hardest known material.